%% file: bola.tex
\documentclass[10pt, journal, letterpaper]{IEEEtran}
\usepackage{graphicx}
\usepackage{amsmath}
\usepackage{amsfonts}
\usepackage{amssymb}
\usepackage{euscript}
\usepackage{comment}
\usepackage{color}
\usepackage{url}
\usepackage{multirow}
\usepackage{placeins}
\usepackage{subfigure}
\usepackage{epsfig}
\usepackage{grffile}
\usepackage{afterpage}
\usepackage{wrapfig}
\usepackage{cite}
\usepackage{setspace}
\usepackage{comment}
\usepackage{enumerate}
\usepackage{float}
\usepackage{algorithmicx}
\usepackage{algpseudocode}
\usepackage[hidelinks=true]{hyperref}

\usepackage{soul}

\usepackage{xcolor}

\newtheorem{thm}{Theorem}

\newtheorem{lem}{Lemma}
\newcommand{\expect}[1]{\mathbb{E}\left\{#1\right\}}
\newcommand{\defequiv}{\mathbin{\mbox{\raisebox{-.3ex}{$\overset{\vartriangle}{=}$}}}}

\DeclareMathOperator*{\argmax}{arg\,max}

\IEEEoverridecommandlockouts

\begin{document}
\title{BOLA: Near-Optimal Bitrate Adaptation \\ for Online Videos
  \thanks{This work was supported in part by the NSF under Grant CNS-1413998 and Grant CNS-1901137. A final version of this paper appears in IEEE/ACM Transactions on Networking 2020 \cite{Spiteri2020}.
}}

\author{
  \IEEEauthorblockN{
    Kevin~Spiteri\textsuperscript{1}\thanks{},
    Rahul~Urgaonkar\,\textsuperscript{2},
    Ramesh~K.~Sitaraman\textsuperscript{1}}

  \IEEEauthorblockA{
    \textsuperscript{1}University of Massachusetts at Amherst, \textsuperscript{2}Amazon Prime Video\\
    \{kspiteri,ramesh\}@cs.umass.edu, urgaonka@amazon.com}
}

\maketitle

\input{abstract}
\input{intro}
\input{theory}

\input{empirical}

\input{deployment}

\input{related}
\input{conclusion}

\section{Acknowledgments}

We would like to thank Daniel Sparacio and Will Law of Akamai for their key insights on real-world player implementations. Further, Daniel was instrumental in helping us implement BOLA in the DASH reference player.

\input{appendix}

\bibliographystyle{IEEEtran}
\bibliography{bola}

\end{document}

%% file: abstract.tex
\begin{abstract}

  Modern video players employ complex algorithms to adapt the bitrate of the video that is shown to the user. Bitrate adaptation requires a tradeoff between reducing the probability that the video freezes (rebuffers) and enhancing the quality of the video. A bitrate that is too high leads to frequent rebuffering, while a bitrate that is too low leads to poor video quality. Video providers segment videos into short segments and encode each segment at multiple bitrates. The video player adaptively chooses the bitrate of each segment to download, possibly choosing different bitrates for successive segments. We formulate bitrate adaptation as a utility-maximization problem and devise an online control algorithm called BOLA that uses Lyapunov optimization to minimize rebuffering and maximize video quality. We prove that BOLA achieves a time-average utility that is within an additive term $O(1/V)$ of the optimal value, for a control parameter V related to the video buffer size. Further, unlike prior work, BOLA does not require prediction of available network bandwidth. We empirically validate BOLA in a simulated network environment using a collection of network traces. We show that BOLA achieves near-optimal utility and in many cases significantly higher utility than current state-of-the-art algorithms. Our work has immediate impact on real-world video players and for the evolving DASH standard for video transmission. We also implemented an updated version of BOLA that is now part of the standard reference player dash.js and is used in production by several video providers such as Akamai, BBC, CBS, and Orange.

\end{abstract}

\begin{IEEEkeywords}
Internet Video, Video Quality, Adaptive Bitrate Streaming, Lyapunov Optimization, Optimal Control
\end{IEEEkeywords}

%% file: intro.tex
\section{Introduction}
\label{section:intro}

Online videos are the ``killer'' application of the Internet with videos currently accounting for more than half of the Internet traffic. Video viewership is growing at a torrid pace and videos are expected to account for more than 85\% of all Internet traffic within a few years \cite{cisco-videogrowth}. As all forms of traditional media migrate to the Internet, video providers face the daunting challenge of providing a good quality of experience (QoE) for users watching their videos. Video providers are diverse and include major media companies (e.g., NBC, CBS), news outlets (e.g., CNN), sports organizations (e.g., NFL, MLB), and video subscription services (e.g., Netflix, Hulu). Recent research has shown that low-performing videos that start slowly, play at lower bitrates, and freeze frequently can cause viewers to abandon the videos or watch fewer minutes of the videos, significantly decreasing the opportunity for generating revenue for the video providers \cite{DobrianSASJGZZ11, KrishnanS2012, sitaraman2013network}, underscoring the need for a high-quality user experience.

Providing a high-quality experience for video users requires balancing two contrasting requirements. The user would like to watch the highest-quality version of the video possible, where video quality can be quantified by the bitrate at which the video is encoded. For instance, watching a movie in high definition (HD) encoded at 2 Mbps arguably provides a better user experience than watching the same movie in standard definition (SD) encoded at a bitrate of 800 kbps. In fact, there is empirical evidence that the user is more engaged and watches longer when the video is presented at a higher bitrate. However, it is not always possible for users to watch videos at the highest encoded bitrate, since the bandwidth available on the network connection between the video player on the user's device and the video server constrains what bitrates can be watched. In fact, choosing a bitrate that is higher than the available network bandwidth\footnote{ Throughout this paper, we say \textbf{bandwidth} when talking about network throughput and \textbf{bitrate} when talking about encoding quality. } will lead to video freezes in the middle of the playback, since the rate at which the video is being played exceeds the rate at which the video can be downloaded. Such video freezes are called \emph{rebuffers} and playing the video continuously \emph{without rebuffers} is a key factor in the QoE perceived by the user \cite{KrishnanS2012}. Thus, balancing the contrasting requirements of playing videos at a high bitrate while at the same time avoiding rebuffers is central to providing a high-quality video watching experience.

\subsection{Adaptive Bitrate (ABR) Streaming}

Achieving a high QoE for video streaming is a major challenge due to the sheer diversity of video-capable devices that include smartphones, tablets, desktops, and televisions. Further, the devices themselves can be connected to the Internet in a multitude of ways, including cable, fiber, DSL, WiFi and mobile wireless, each providing different bandwidth characteristics. The need to adjust the video playback to the characteristics of the device and the network has led to the evolution of adaptive bitrate (ABR) streaming that is now the de facto standard for delivering videos on the Internet.

ABR streaming requires that each video is partitioned into \emph{segments,} where each segment corresponds to a few seconds of play. Each segment is then encoded in a number of different bitrates to accommodate a range of device types and network connectivities. When the user plays a video, the video player can download each segment at a bitrate that is appropriate for the available bandwidth of the network connection. Thus, the player can switch to a segment with a lower bitrate when the available bandwidth is low to avoid rebuffering. If more bandwidth becomes available at a future time, the player can switch back to a higher bitrate to provide a richer experience. The video player has a buffer that allows it to fetch and store segments \emph{before} they need to be rendered on the screen. Thus, the video player can tolerate brief network disruptions without interrupting the playback of the user by using the buffered segments. A large disruption, however, will empty the buffer, resulting in rebuffering. The decision of which segments to download at what bitrates is made by a \emph{bitrate adaptation algorithm} within the video player, the design of such algorithms being the primary focus of our work.

Several popular implementations of ABR streaming exist, including Apple's HTTP Live Streaming (HLS)~\cite{apple}, Microsoft's Live Smooth Streaming (Smooth)~\cite{smooth} and Adobe's Adaptive Streaming (HDS)~\cite{adobe}. Each has its own proprietary implementation and slight modifications to the basic ABR technique described above. A key recent development is a unifying open-source standard for ABR streaming called MPEG-DASH~\cite{Stockhammer2011}. DASH is broadly similar to the other ABR protocols and is a particular focus in our empirical evaluation.

\subsection{Our Contributions}

Our primary contribution is a principled approach to the design of bitrate adaptation algorithms for ABR streaming. In particular, we formulate bitrate adaptation as a utility maximization problem that incorporates both key components of QoE: the average bitrate of the video experienced by the user and the duration of the rebuffer events. An increase in the average bitrate increases utility, whereas rebuffering decreases it. A strength of our framework is that utility can be defined in a very general manner, say, depending on the content, video provider, or user device.

Using Lyapunov optimization, we derive an online bitrate adaptation algorithm called \mbox{BOLA} (Buffer Occupancy based Lyapunov Algorithm) that provably achieves utility that is within an additive factor of the maximum possible utility in the large video regime. While numerous bitrate adaptation algorithms have been proposed \cite{jiang2014festive,Huang+14,decicco2013elastic,li2014panda,yin2015ccontrol,mao2017neural} and implemented within video players, our algorithm is the first to provide a \emph{theoretical guarantee} on the achieved utility. Further, \mbox{BOLA} provides an explicit knob for video providers to set the relative importance of a high video quality in relation to the probability of rebuffering.

While not an explicit part of the Lyapunov optimization framework, we also show how \mbox{BOLA} can be adapted to avoid frequent bitrate switches during video playback. Bitrate switches are arguably less annoying than rebuffering, but it is still of some concern to video providers and users alike if such switches occur too frequently.

Most algorithms implemented in practice use a \emph{bandwidth-based} approach where the available bandwidth between the server and the video player is predicted and the predicted value is used to determine the bitrate of the next segment that is to be downloaded. A complementary approach is a \emph{buffer-based} approach that does not predict the bandwidth, but only uses the amount of data that is currently stored in the buffer of the video player. Recently, there has been empirical evidence that a buffer-based approach has desirable properties that bandwidth-based approaches lack and has been adopted by Netflix \cite{Huang+14}. An intriguing outcome of our work is that the optimal algorithm within our utility maximization framework requires only knowledge of the amount of data in the buffer and no estimate of the available bandwidth. Thus, our work provides the first theoretical justification for why buffer-based algorithms perform well in practice and adds new insights to the ongoing debate \cite{yin2015ccontrol} within the video streaming and DASH standards communities of relative efficacy of the two approaches. Further, since our algorithm BOLA is buffer-based, it avoids the overheads of more complex bandwidth prediction present in current video player implementations and is more stable under bandwidth fluctuations. Note that our results imply that the buffer level is a \emph{sufficient statistic} that indirectly provides all information about past bandwidth variations required for choosing the next bitrate.

We also \emph{empirically} evaluate \mbox{BOLA} on a wide set of network traces that include 12 test cases provided by the DASH industry forum \cite{dashtest} and 85 publicly-available 3G mobile bandwidth traces \cite{riiser2013}. As a benchmark for comparison, we develop an optimal \emph{offline} algorithm that uses dynamic programming and is guaranteed to produce the maximum achievable time-average utility for any given set of network traces. Unlike \mbox{BOLA} that works in an online fashion, the offline optimal algorithm makes decision based on perfect knowledge of future bandwidth variations. Remarkably, the utility achieved by \mbox{BOLA} is within 84--95\% of offline optimal utility for all the tested traces.

Besides comparing \mbox{BOLA} with the offline optimal, we also empirically compared our algorithm with four state-of-the-art algorithms proposed in the literature. In \emph{all} test cases, \mbox{BOLA} achieved a utility that is as good as or better than the best state-of-the-art algorithm.

We also implemented BOLA in dash.js, the open-source standard DASH reference player \cite{dashjs}. Deploying BOLA in production required a number of adjustments \cite{Spiteri2018}. Through dash.js, BOLA is now being used in production by several major video providers and delivery networks such as Akamai, BBC, CBS and Orange. BOLA is available as an option to commercial video providers who often use the production dash.js reference implementation for building their own video players. Further, a second algorithm called DYNAMIC \cite{Spiteri2018} is also available for commercial video providers. DYNAMIC is a hybrid algorithm that uses a simple throughput-estimation approach during the start-up phase of the video and then uses BOLA afterwards. Both algorithms can be evaluated in a web browser by clicking the ``Show Options'' button in the latest version of the dash.js reference player found at \cite{dashjs}.

%% file: theory.tex
\section{System Model}
\label{sec:model}

Our system model closely captures how ABR streaming works on the Internet today. We consider a video player that downloads a video file from a server over the Internet and plays it back to the user. The video file is segmented into segments that are downloaded in succession. The available bandwidth between the server and the player varies over time. This can be due to reasons such as network congestion and wireless fading among others. The viewing experience of the user is determined by both the video quality as quantified by the bitrates of the segments that are played back and the playback characteristics such as rebuffering. The objective of the player is to maximize a utility associated with the user's viewing experience while adapting to time-varying (and possibly unpredictable) changes in the available bandwidth.

\emph{Video Model:}
The video file is segmented into $N$ segments indexed as $\{1, 2, \ldots, N\}$ where each segment represents $p$ seconds of the video. On the server, each segment is available in $M$ different bitrates where a segment encoded at a higher bitrate has a larger size in bits and its playback provides a better user experience and higher utility. Suppose the size (in bits) of any\footnote{For simplicity, we assume that the segment size (in bits) is $S_m$ for all segments of a given bitrate index $m$. However, our framework can be easily extended to the case where the segment size for the same bitrate can vary across segments.} segment encoded at bitrate index $m$ is $S_m$ bits and suppose the utility derived by the user from viewing it is given by $\upsilon_m$ where $m \in \{1, 2, \ldots, M\}$. WLOG, let the segment bitrates be non-decreasing in index $m$. Then, the following holds.
\begin{align}
\upsilon_1 \leq \upsilon_2 \leq \ldots \leq \upsilon_M \Longleftrightarrow S_1 \leq S_2 \leq \ldots \leq S_M.
\label{eq:upsilon}
\end{align}
Note that the actual encoding bitrate for bitrate index $m$ is given by $S_m/p$ bits/second.

\emph{Video Player:} The video player downloads successive segments of the video file from the server and plays back the downloaded segments to the user. Each segment must be downloaded in its entirety before it can be played back. We assume that the player sends requests to the server to download one segment at a time. Also, the segments are downloaded in the same order as they are played back.	The video player has a finite buffer of size $Q_{\max}$ segments\footnote{It is common practice for video players to measure the buffer in seconds of playback time rather than in bits.} to store the downloaded but yet-to-be-played-back segments. Measuring the buffer in segments is equivalent to measuring it in seconds since the segment duration $p$ is fixed. If the buffer is full the player cannot download any new segments and waits for a fixed period of time given by $\Delta$ seconds before attempting to download a new segment. The segments that are fully downloaded are played back at a fixed rate of $1/p$ segments/second without any idling.

When sending a download request for a new segment, the player also specifies the desired bitrate for that segment. This enables the player to tradeoff the overall video quality with the likelihood of rebuffering that occurs when there are no segments in the buffer for playback. Note that while each segment has a fixed playback time of $p$ seconds, the size of the segment (in bits) can be different depending on its bitrate. Thus, the choice of bitrate for a segment impacts its download time.

\emph{Network Model:}
The available bandwidth (in bits/second) between the server and player is assumed to vary continuously in time according to a stationary random process $\omega(t)$. We do not make any assumptions about knowing the statistical properties or probability distribution of $\omega(t)$ except that it has finite first and second moments as well as a finite inverse second moment. Suppose the player starts to download a segment of bitrate index $m$ at time $t$. Then the time $t'$ when the download finishes satisfies the following:

\begin{align}
S_m = \int_t^{t'} \omega(\tau) d\tau
\end{align}
Let $\expect{\omega(t)} = \omega_{\mathrm{avg}}$. Then, $\expect{t' - t} = {S_m}/{\omega_{\mathrm{avg}}}$.

\section{Problem Formulation}
\label{sec:formulation}

We consider two primary performance metrics\footnote{We do not include the secondary objective of avoiding frequent bitrate switches in our formulation, but we deal with it empirically in Section~\ref{subsec:oscillations}.} that affect the overall QoE of the user: (1) time-average playback quality which is a function of the bitrates of the segments viewed by the user 
and (2) fraction of time spent not rebuffering. To formalize these metrics, we consider a time-slotted representation of our system model. The timeline is divided into non-overlapping consecutive slots of variable length and indexed by $k \in \{1, 2, \ldots\}$. Slot $k$ starts at time $t_k$ and is $T_k = t_{k+1} - t_k$ seconds long. We assume that $t_1 = 0$. At the 
beginning of each slot, the video player makes a control decision on whether it should start downloading a new segment, and if yes, its bitrate. If a download decision is made, then a request is sent to the server and the download starts immediately\footnote{Any delays associated with sending the request can be added to the overall download time.}. This download takes 
$T_k$ seconds and is completed at the end of slot $k$. Note that $T_k$ is a random variable whose actual value depends on the realization of the $\omega(t)$ process as well as the choice of segment bitrate. If the player decides not to download a new segment in slot $k$ (for example, when the buffer is full), then this slot lasts for a fixed duration of $\Delta$ 
seconds.

We define the following indicator variable for each slot $k$:
\begin{equation}
a_{m}(t_k) = \left\{ \begin{array}{ll}
1  & \textrm{if the player downloads a segment }\\
   & \textrm{of bitrate index $m$ in slot $k$, and}\\ 0 & \textrm{otherwise.} \end{array} \right. 
\label{eq:a_m} 
\end{equation} 
Then, for all $k$, we must have $\sum_{m=1}^M a_{m}(t_k) \leq 1$. Moreover, when $\sum_{m=1}^M a_{m}(t_k) = 0$, then no segments are downloaded.

Denote the buffer level (measured in number of segments) at the start of slot $k$ by $Q(t_k)$. The dynamics of this queue can be expressed using the following equation:
\begin{align}
Q(t_{k+1}) = \max[Q(t_k) - \frac{T_k}{p}, 0] + \sum_{m=1}^M a_m(t_k)
\label{eq:Q}
\end{align}
Here, the arrival value into this queue in slot $k$ is given by $\sum_{m=1}^M a_m(t_k)$ which is $1$ if a download decision is made in slot $k$ and $0$ otherwise. 
The departure value is $T_k/p$ which represents the total number of segments (including fractional segments) that could have departed the buffer in slot $k$. 
Note that the actual value of $T_k$ is revealed at the end of slot $k$. 
Also note that a segment that is downloaded in slot $k$ becomes available for playback only from the next slot. 
We assume that the buffer level is initialized to $0$, i.e., $Q(t_1) = 0$.

Let $K_N$ denote the index of the slot in which the $N^{\textrm{th}}$ (i.e., last) segment is 
downloaded. Also, denote the time at which the player finishes playing back the last segment by $T_{\mathrm{end}}$. Then the first performance metric of interest is the time-average expected {\em playback utility} $\overline{\upsilon}_N$ which is defined as
\begin{align}
\overline{\upsilon}_N \defequiv \frac{\expect{ \sum_{k=1}^{K_N} \sum_{m=1}^M a_{m}(t_k) \upsilon_m}} {\expect{T_{\mathrm{end}}}}
\end{align}
where the numerator denotes the expected total utility across all $N$ segments. Note that a segment can only be played back after it has been downloaded entirely. Thus, $T_{\mathrm{end}}$ is greater than the last segment's download finish time, i.e., $T_{\mathrm{end}} > t_{K_N} + T_{K_N}$.

The second performance metric of interest is the expected fraction of time $\overline{s}_N$ that is spent not rebuffering and can be interpreted as a measure of the average playback ``smoothness''. This can be calculated by observing that the actual playback time for all $N$ segments is $Np$ seconds. Thus, the expected {\em playback smoothness} $\overline{s}_N$ is 
given by
\begin{align}
\overline{s}_N \defequiv \frac{Np} {\expect{T_{\mathrm{end}}}} = \frac{\expect{ \sum_{k=1}^{K_N} \sum_{m=1}^M a_{m}(t_k) p}} {\expect{T_{\mathrm{end}}}}
\end{align}
where in the last step we use the relation that $Np = \sum_{k=1}^{K_N} \sum_{m=1}^M a_{m}(t_k) p$. Note that $T_{\mathrm{end}} \geq Np$ (since at most one segment can be played back at any time), so that $\overline{s}_N \leq 1$.

\subsection{Design Objective}
\label{section:objective}

We want to design a control algorithm that maximizes the joint utility $\overline{\upsilon}_N + \gamma \overline{s}_N$ subject to the constraint that $Q(t_k) \leq Q_{\max}$ for all $k$. Here, $\gamma > 0$ is an input weight parameter for prioritizing playback utility versus the playback smoothness.

This problem can be formulated as a stochastic optimization problem with a time-average objective over a finite horizon and dynamic programming (DP) based approaches can be used to solve it \cite{bertsekas1995dynamic}. However, traditional DP based methods have two major disadvantages. First, they require knowledge of the distribution of the $\omega(t)$ process which 
may be hard to obtain. Second, even when such knowledge is available, the resulting DP can have a very large state space. This is because the state space for this problem under a DP formulation would consist of not only the timeslot index $k$ and value $t_k$, but also the buffer size $Q(t_k)$. Further, an appropriate discretization of the $\omega(t)$ process would be 
required to obtain a tractable solution.

In order to overcome the above challenges associated with traditional DP based methods, we take an alternate approach in this paper. First, we consider the bitrate adaptation problem in the limiting regime when the video size becomes large, i.e., $N \to \infty$. 
Second, we replace the finite buffer constraint with a \emph{rate stability} constraint (made precise in 
the next section). The reason for making these assumptions is that it results in simplifications to the original problem as discussed in the next section. This allows us to develop a bitrate adaptation algorithm that does not require \emph{any} knowledge of the distribution of $\omega(t)$, yet offers provable theoretical performance guarantees in the large video size regime 
while satisfying the finite buffer constraint. As shown later in Section~\ref{subsec:bolaFinite}, with slight modifications, this algorithm can be used for finite sized videos as well and offers close to optimal performance in our experiments.



\subsection{Problem Relaxation}
\label{section:relax}

Consider the bitrate adaptation problem in the limiting regime when the video size becomes large, i.e., $N \to \infty$. Then, the metrics $\overline{\upsilon}_N$ and $\overline{s}_N$ can be expressed as

\begin{align}
\overline{\upsilon}  \defequiv \lim_{N \to \infty}\! \overline{\upsilon}_N  &= \lim_{N \to \infty}  \frac{\expect{ \sum_{k=1}^{K_N} \sum_{m=1}^M a_{m}(t_k) \upsilon_m}} {\expect{T_{\mathrm{end}}}} \nonumber \\
& = \frac{ {\displaystyle \lim_{K_N \to \infty}} \frac{1}{K_N}  \expect{\! \sum_{k=1}^{K_N}\! \sum_{m=1}^M\! a_{m}(t_k) \upsilon_m \!}} {{\displaystyle \lim_{K_N \to \infty}} \frac{1}{K_N} \expect{ \sum_{k=1}^{K_N} T_{k}}} \label{eq:inf_u} \\
 \overline{s}  \defequiv  \lim_{N \to \infty} \overline{s}_N  &= \lim_{N \to \infty} \frac{\expect{ \sum_{k=1}^{K_N} \sum_{m=1}^M a_{m}(t_k) p}} {\expect{T_{\mathrm{end}}}} \nonumber\\ 
& = \frac{{\displaystyle \lim_{K_N \to \infty}} \frac{1}{K_N} \expect{ \sum_{k=1}^{K_N}  \sum_{m=1}^M a_{m}(t_k) p}} {{\displaystyle \lim_{K_N \to \infty}} \frac{1}{K_N} \expect{\sum_{k=1}^{K_N}  T_{k}}} \label{eq:inf_s}
\end{align}
This follows by noting that the difference between the expected total playback finish time $\expect{T_{\mathrm{end}}}$ and the expected total download finish time $\expect{\sum_{k=1}^{K_N} {T_{k}}}$ is upper bounded by a finite value due to the finite $Q_{\max}$. Specifically, this upper bound is given by $Q_{\max} p$. Therefore, instead of considering the total 
playback finish time, we can consider the total download finish time in the objective when the video size becomes large.

Next, replace the finite buffer constraint with a rate stability constraint \cite{neely2010}. This constraint only requires that the time-average arrival rate into the buffer cannot exceed the time-average playback rate. This is equivalent to requiring that
\begin{align}
{{\displaystyle \lim_{K_N \to \infty}} \frac{1}{K_N} \expect{ \sum_{k=1}^{K_N}  \sum_{m=1}^M a_{m}(t_k) p}} \leq {{\displaystyle \lim_{K_N \to \infty}} \frac{1}{K_N} \expect{\sum_{k=1}^{K_N}  T_{k}}} 
\label{eq:rate_stability}
\end{align}
The rate stability constraint is a relaxation of the finite buffer constraint since any policy that ensures finite buffers is always rate stable but not vice versa. 
Therefore, under this relaxation, the optimal time-average utility cannot be smaller than the optimal time-average utility with the finite buffer constraint.

With these relaxations, our performance objective for the bitrate adaptation problem is to maximize the joint utility $\overline{\upsilon} + \gamma \overline{s}$ subject to 
the rate stability constraint \eqref{eq:rate_stability}.%
 Let us denote the optimal time-average utility for this problem by $\upsilon^* + \gamma s^*$. 
This problem fits in the framework of Lyapunov optimization for renewal systems \cite{neely2012dynamic}.%
 Specifically, this framework extends the original Lyapunov optimization technique \cite{neely2010} to systems with variable length renewal frames and shows that  
minimizing a ``drift-plus-penalty'' ratio over every frame yields an optimal control algorithm.
We refer to \cite{neely2012dynamic} for details on this method. 
In the context of our bitrate adaptation problem, the variable length slots represent the renewal frames.

The following characterization can be made about the optimality of \emph{i.i.d. algorithms}.
\begin{lem} 
\label{lem:policy} 
For the bitrate adaptation problem in the limiting regime when the video size becomes large, i.e., $N \to \infty$, there exists a 
\emph{buffer-state-independent} stationary algorithm that makes i.i.d. control decisions in every slot and satisfies the rate stability constraint while achieving 
time-average utility no smaller than $\upsilon^* + \gamma s^*$.
\end{lem}
\begin{IEEEproof}
This follows from Lemma $1$ in \cite{neely2012dynamic} and uses the fact that the conditional expectations and conditional second moments of the frame length and utility are bounded under any algorithm. The full proof is omitted for brevity.
\end{IEEEproof}

Note that such a {buffer-state-independent} stationary algorithm is not necessarily feasible for our finite buffer system. Further, calculating it explicitly would require 
knowledge of the distribution of $\omega(t)$. However, instead of calculating this policy explicitly, we will use its existence and characterization per 
Lemma~\ref{lem:policy} to design an \emph{online} control algorithm using the technique of Lyapunov optimization over renewal frames.

In the next section, we will present this algorithm and show that it meets the finite buffer constraint while achieving a time-average utility that is within $O(1/Q_{\max})$ of $\upsilon^* + \gamma s^*$ without requiring any knowledge of the distribution of $\omega(t)$.

\section{BOLA: An Online Control Algorithm}
\label{section:online_algo}

We first give a high-level intuition of the Lyapunov optimization over renewals technique. This technique converts the problem of optimizing the time-average metrics
in \eqref{eq:inf_u}--\eqref{eq:inf_s} subject to the time-average constraint in \eqref{eq:rate_stability} into a series of per slot optimization problems. The problem 
to be solved in each slot involves minimizing a ratio of the expected drift-plus-penalty value in that slot to the expected length of the slot. 
As shown in the Appendix, this can be done without requiring any knowledge of  the distribution of $\omega(t)$.
The drift term consists of $\expect{(Q(t_{k+1})^2-Q(t_k)^2)/2 \mid Q(t_k)}$ and serves to meet the rate stability constraint \eqref{eq:rate_stability}. The penalty term consists of the playback utility and playback smoothness received in that slot. We keep the utility and smoothness as separate terms even though they can be folded into one metric. This allows us to  tune the relative importance of increasing video bitrate and reducing rebuffering without changing the algorithm.
The algorithm uses a control parameter $V > 0$ to allow a tradeoff between the buffer size and the performance objectives.

We now present the algorithm. In every slot $k$, given the buffer level $Q(t_k)$ at the start of the slot, our algorithm makes a control decision by solving the following deterministic optimization problem. Let
\begin{align}
  \rho(t_k, \mathbf{a}(t_k)) &= \begin{cases}
    0 \hfill \text{if $\sum_{m=1}^M  a_m(t_k) = 0$}, \vspace{12pt} \\
    \dfrac{\sum_{m=1}^M  a_m(t_k) \big(V\upsilon_m + V\gamma p - Q(t_k) \big)}{\sum_{m=1}^M  a_m(t_k) S_m} \\ \hfill \text{otherwise.}
  \end{cases}
  \label{eq:lyp_score}
\end{align}

Then determine $\mathbf{a}(t_k)$ by solving the optimization problem:
\begin{align}
  & \textrm{Maximize:}  && \rho(t_k, \mathbf{a}(t_k)) \nonumber \\
  & \textrm{Subject to:}  && \textstyle\sum_{m=1}^M a_{m}(t_k) \leq 1, a_{m}(t_k) \in \{0, 1\}
\label{eq:lyp_algo}
\end{align}

The constraints of this problem result in a very simple solution structure. Specifically, the optimal solution is given by:
\begin{enumerate}
\item
If $Q(t_k) > V ( \upsilon_m + \gamma p)$ for all $m \in \{1, 2, \ldots, M\}$, then the no-download option is chosen, i.e., $a_m(t_k) = 0$ for all $m$. Note that in this case $T_k = \Delta$.
\item
Else, the optimal solution is to download the next segment at bitrate index $m^*$ where $m^*$ is the index that maximizes the ratio $\big(V\upsilon_m + V \gamma p - Q(t_k)\big)/{S_m}$ among all $m$ for which this ratio is positive.
\end{enumerate}

Notice that solving this problem does not require any knowledge of the $\omega(t)$ process. Further, the optimal solution depends only on the buffer level $Q(t_k)$. That's why we call our algorithm \emph{BOLA: Buffer Occupancy based Lyapunov Algorithm}. These properties of \mbox{BOLA} should be contrasted with the bandwidth prediction based strategies that have been recently proposed for this problem that require explicit prediction of the available bandwidth for control decisions.

The following theorem characterizes the theoretical performance guarantees provided by \mbox{BOLA}.
\begin{thm}
\label{thm:BOLA}
Suppose \mbox{BOLA} as defined by (\ref{eq:lyp_algo}) is implemented in every slot using a control parameter $0 < V \leq \frac{Q_{\max} - 1}{\upsilon_M + \gamma p}$. Assume $Q(0) = 0$. Then, the following hold.
\begin{enumerate}
\item
The queue backlog satisfies $Q(t_k) \leq V(\upsilon_M + \gamma p) + 1$ for all slots $k$. Further, the buffer occupancy in segments never exceeds $Q_{\max}$.
\item
The time-average utility achieved by \mbox{BOLA} satisfies
\begin{align}
\overline{\upsilon}^{\mathrm{BOLA}} + \gamma \overline{s}^{\mathrm{BOLA}} \geq  \upsilon^* + \gamma s^* - \frac{p^2 + \Psi}{2p^2V}
\label{eq:bound2}
\end{align}
where $\Psi$ is an upper bound on $\expect{T_k^2}$ under any control algorithm and is assumed to be finite.
\end{enumerate}
\label{thm:perf_bounds}
\end{thm}
\begin{IEEEproof}
See the \hyperref[sec:NP-proof]{Appendix}.
\end{IEEEproof}
\emph{Remarks:}
The performance bounds in Theorem \ref{thm:perf_bounds} show a $[O(1/V), O(V)]$ utility and backlog tradeoff that is typical of Lyapunov based control algorithms for similar utility maximization problems. Specifically, the time-average utility of \mbox{BOLA} is within an $O(1/V)$ additive term of the optimal utility and this gap may be made smaller by  choosing a larger value of $V$. However, the largest feasible value of $V$ is constrained by the buffer size and there is a linear relation between them.

\subsection{Understanding BOLA With an Example}
\label{subsec:example}

We now present a sample run to illustrate how BOLA works. We slice a 99-second video using 3-second segments and encode it at five different bitrates. While BOLA only requires the utilities to be a non-decreasing function of the segment bitrate, it is natural to consider concave utility functions with diminishing returns, e.g., a 1 Mbps increase in segment bitrate likely provides a larger utility gain for the user when that increase is from 0.5~Mbps to 1.5~Mbps than when it is from 5~Mbps to 6 Mbps. A natural choice for our example is the logarithmic utility function: let $\upsilon_m = \ln(S_m / S_1)$. Pick $\gamma=5.0 / p$ and $V=0.93$. The bitrates and utilities are below.

\begin{center}
  \scriptsize
  \begin{tabular}{|c||c|c|c|c|c|}
    \hline
    bitrate (Mbps) & 0.331 & 0.688 & 1.427 & 2.962 & 6.000 \\
    \hline
    $S$ (Mb)       & 0.993 & 2.064 & 4.281 & 8.886 & 18.00 \\
    \hline
    $\upsilon$     & 0.000 & 0.732 & 1.461 & 2.192 & 2.897 \\
    \hline
  \end{tabular}

\end{center}

\begin{figure}
  \centering
  \includegraphics{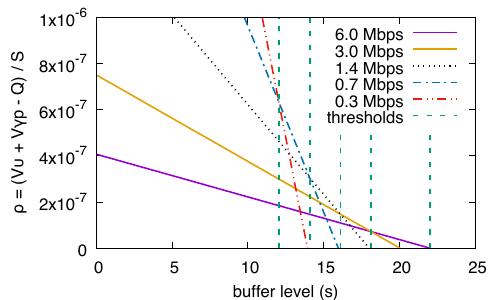}
  \caption{The value of $(V \upsilon_m + V \gamma p - Q) / S_m$ for different bitrates depends on buffer level. ($\gamma p = 5$ and $V = 0.93$.) Note that the buffer level is $Qp$ seconds.}
  \label{fig:DemoThreshold}
\end{figure}

\begin{figure}
  \centering
  \includegraphics{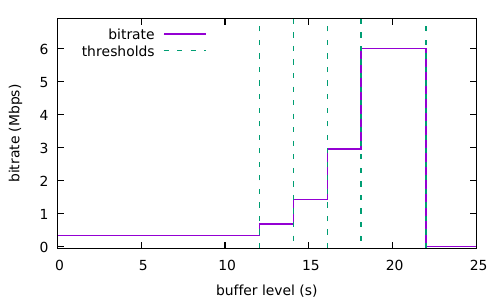}
  \caption{\mbox{BOLA}'s bitrate choice as function of buffer level. ($\gamma p = 5$,$V = 0.93$.) Note that the buffer level is $Qp$ seconds.}
  \label{fig:DemoDecision}
\end{figure}

\begin{figure}
  \centering
  \includegraphics{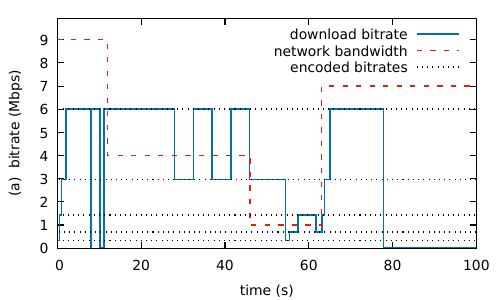} \\
  \includegraphics{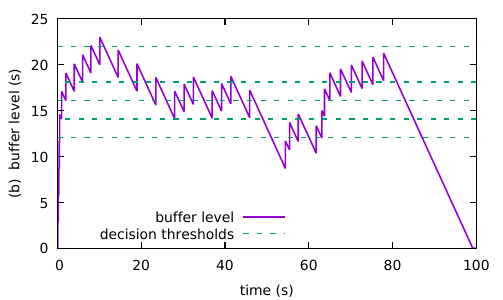}
  \caption{Sample video download and playback using \mbox{BOLA}. (a) The video is encoded at 5 different bitrates. The network bandwidth varies from high to low and back to high. The downloaded segment bitrate adapts to the network bandwidth. (b) The buffer level variation triggers bitrate changes when it crosses the thresholds.}
  \label{fig:Demo}
\end{figure}

For any slot we choose the segment bitrate to maximize $(V \upsilon_m + V \gamma p - Q) / S_m$ for $1 \leq m \leq M$. Fig.~\ref{fig:DemoThreshold} shows the relationship between the expression and the buffer level $Q$ for different $m$. The line intersections mark the buffer levels that correspond to decision thresholds. Fig.~\ref{fig:DemoDecision} summarizes \mbox{BOLA}'s bitrate choices as a function of the buffer level. 

Fig.~\ref{fig:Demo} shows how \mbox{BOLA} works. We use a synthetic network bandwidth profile as shown in Fig.~\ref{fig:Demo}(a). We can see the feedback loop involving the bitrate in (a) and the buffer level in (b). \mbox{BOLA} chooses the bitrate based directly on the buffer level using Fig.~\ref{fig:DemoDecision}. The bitrate affects the download time, thus it indirectly affects the buffer level at the beginning of the following slot. Finally, when all the segments are downloaded, the video player plays out the segments remaining in the buffer.

\subsection{\texorpdfstring{Choosing Utility and Parameters $\gamma$ and $V$}{Choosing Utility and Parameters g and V}}
\label{subsec:param}

While we chose a logarithmic utility function for the example, a video provider can use any utility function satisfying~\eqref{eq:upsilon}. The utility function might also take into account system characteristics such as the type of device a viewer is using.

$\gamma$ corresponds to how strongly we want to avoid rebuffering. Increasing $\gamma$ translates the graphs in Figs.~\ref{fig:DemoThreshold} and~\ref{fig:DemoDecision} to the right, effectively shifting the thresholds higher without changing their relative distance. \mbox{BOLA} will thus download more low-bitrate segments to maintain a larger (and safer) buffer level.

Increasing $V$ expands the graphs in Figs.~\ref{fig:DemoThreshold} and~\ref{fig:DemoDecision} horizontally about the origin. If we have a maximum buffer level $Q_{\max}$ we want to avoid downloading unless there is enough space for one full segment on the buffer, that is unless $Q \leq Q_{\max} - 1$. For a given $Q_{\max}$ we can set $V = (Q_{\max} - 1) / (\upsilon_M + \gamma p)$.

While we showed how to choose reasonable values for $\gamma$ and $V$, video providers are more familiar with choosing buffer level targets. A method to derive the parameters from buffer level targets is included in Section~\ref{sec:deployparam}.
Alternatively, video providers might choose $\gamma$ and $V$ by employing an approach such as Oboe~\cite{akhtar2018oboe} to auto-tune the BOLA parameters.

%% file: empirical.tex
\section{Implementation and Empirical Evaluation}

\label{sec:empirical}

We first implemented a basic version of \mbox{BOLA}, named \mbox{BOLA-BASIC}, directly from \eqref{eq:lyp_algo}. Recall that when the buffer level is full \mbox{BOLA} does not download a segment but waits for $\Delta$~seconds. Rather than picking an arbitrary value for $\Delta$, we use a dynamic wait until $Q(t_k) \leq V(\upsilon_M + \gamma p)$. This has the same effect as picking a fixed but very small $\Delta$, so the theoretical analysis still holds. We also implemented other versions of \mbox{BOLA}, namely \mbox{BOLA-FINITE}, \mbox{BOLA-O}, and \mbox{BOLA-U}, that we describe later in this section.

\subsection{Test Methodology}

We simulated all versions of \mbox{BOLA} using the Big Buck Bunny movie \cite{bigbuckbunny}. The 10-minute movie was encoded at 10 different bitrates and sliced in 3-second segments. Although each quality index has a specified average bitrate, segments may have variable bitrate (VBR) because of the varying nature of the movie. We simulate playback times longer than 10 minutes by repeating the movie. Again we choose a logarithmic utility function: $\upsilon_m = \ln(S_m / S_1)$. Table~\ref{tab:bitrates} shows the mean and standard deviation of the bitrate and segment size for each quality index and the respective utility values.

\begin{table}
  \caption{Bitrates used for Big Buck Bunny Test Video}
  \label{tab:bitrates}
  \centering
  \scriptsize
  \begin{tabular}{r|cc|cc|c}
    \hline
    \multicolumn{1}{c|}{Bitrate} &
    \multicolumn{2}{c|}{Bitrate (Mbps)} &
    \multicolumn{2}{c|}{Segment Size $S$ (Mb)} &
    Utility \\
    \multicolumn{1}{c|}{Index} &
    Mean & Standard & Mean   & Standard & $\upsilon$ \\
    \multicolumn{1}{c|}{$m$} &
    & Deviation & & Deviation & ${}=\ln(S/S_1)$ \\
    \hline
    1        & 0.230   & 0.038    & 0.690 & 0.113    &  0.000   \\
    2        & 0.331   & 0.054    & 0.993 & 0.162    &  0.364   \\
    3        & 0.477   & 0.096    & 1.431 & 0.287    &  0.729   \\
    4        & 0.688   & 0.120    & 2.064 & 0.360    &  1.096   \\
    5        & 0.991   & 0.182    & 2.973 & 0.545    &  1.461   \\
    6        & 1.427   & 0.275    & 4.281 & 0.825    &  1.825   \\
    7        & 2.056   & 0.394    & 6.168 & 1.182    &  2.190   \\
    8        & 2.962   & 0.564    & 8.886 & 1.691    &  2.556   \\
    9        & 5.027   & 0.891    & 15.08 & 2.673    &  3.084   \\
    $M={}$10 & 6.000   & 1.078    & 18.00 & 3.232    &  3.261   \\
    \hline
  \end{tabular}
\end{table}

\begin{table}
  \caption{Network Profiles for the Dash Benchmarks}
  \label{tab:profiles}
  \centering
  \scriptsize
  \begin{tabular}{r@{~(}r@{)}r|r@{~(}r@{)}r|r@{~(}r@{)}r||r@{~(}r@{)}r|r@{~(}r@{)}r|r@{~(}r@{)}r}
    \hline
    \multicolumn{3}{c|}{1} & \multicolumn{3}{c|}{3} & \multicolumn{3}{c||}{5} & \multicolumn{3}{c|}{7} & \multicolumn{3}{c|}{9} & \multicolumn{3}{c}{11} \\
    \multicolumn{3}{c|}{Mbps\,(ms)\!} & \multicolumn{3}{c|}{Mbps\,(ms)\!} & \multicolumn{3}{c||}{Mbps\,(ms)\!} & \multicolumn{3}{c|}{Mbps\,(ms)\!} & \multicolumn{3}{c|}{Mbps\,(ms)\!} & \multicolumn{3}{c}{Mbps\,(ms)\!} \\
    \hline
    5.0 &  38 && 5.0 &  13 && 5.0 &  11 &&  \multicolumn{2}{}{} && \multicolumn{2}{}{} && \multicolumn{2}{}{} \\
    4.0 &  50 && 4.0 &  18 && 4.0 &  13 &&  9.0 &  25 && 9.0 &  10 && 9.0 &   6 \\
    3.0 &  75 && 3.0 &  28 && 3.0 &  15 &&  4.0 &  50 && 4.0 &  50 && 4.0 &  13 \\
    2.0 &  88 && 2.0 &  58 && 2.0 &  20 &&  2.0 &  75 && 2.0 & 150 && 2.0 &  20 \\
    1.5 & 100 && 1.5 & 200 && 1.5 &  25 &&  1.0 & 100 && 1.0 & 200 && 1.0 &  25 \\
    2.0 &  88 && 2.0 &  58 && 2.0 &  20 &&  2.0 &  75 && 2.0 & 150 && 2.0 &  20 \\
    3.0 &  75 && 3.0 &  28 && 3.0 &  15 &&  4.0 &  50 && 4.0 &  50 && 4.0 &  13 \\
    4.0 &  50 && 4.0 &  18 && 4.0 &  13 &&  \multicolumn{2}{}{} && \multicolumn{2}{}{} && \multicolumn{2}{}{} \\
    \hline
  \end{tabular}
\end{table}

The DASH Industry Forum provides benchmarks for various aspects of the DASH standard \cite{dashtest}. The benchmarks include twelve different network profiles. Profiles 1--6 have network bandwidths ranging from 1.5 to 5~Mbps while profiles 7--12 have bandwidths ranging from 1 to 9~Mbps. Different latencies are provided for each bandwidth, where the latency is half the round-trip time (RTT). Table~\ref{tab:profiles} shows the odd-numbered bandwidth characteristics. Profile 1 spends 30s at each of 5, 4, 3, 2, 1.5, 2, 3 and 4~Mbps respectively, then starts back at the top. Even-numbered profiles are similar to the preceding odd-numbered profiles but start at the low bandwidth stage. For example, profile~2 starts at 1.5~Mbps.

In addition, we also tested our algorithms using a set of 86 3G mobile bandwidth traces that are publicly available \cite{riiser2013}. One trace was excluded because it had an average bandwidth of 80~kbps; our lowest video bitrate is 230~kbps. Since the traces do not include latency measurements, we used 50~ms latency giving a RTT of 100~ms throughout. This is the median RTT measured empirically in \cite{romirer2009network}.

\subsection{Computing an Upper Bound on the Maximum Utility}

\label{subsec:bound}

In order to evaluate how well \mbox{BOLA} performs on the traces, it is important to derive an upper bound on the maximum utility that is obtainable by {\em any} algorithm on a given trace. We derive an {\em offline} optimal algorithm that provides the maximum achievable utility using dynamic programming. We define a table $r(n,t,b)$ that contains the maximum utility possible when we download the $n^{\text{th}}$ segment and finish at time $t$ with buffer level $b$. We initialize the table with $r(0,0,0) = 0$. Let $x(n,t,m)$ be the time to download the $n^{\text{th}}$ segment at bitrate index $m$ starting at time $t$. Note that the dependency of $x$ on $n$ is due to VBR. We quantize the time with granularity $\delta$. While some accuracy is lost, we ensure the final result will still be an upper bound by rounding the download time down.
\begin{equation*}
  x_{\delta}(n, t, m) = \lfloor x(n, t, m) / \delta \rfloor \cdot \delta
\end{equation*}
We cap the buffer level at $b_{\max}$.
\begin{equation*}
  x'_{\delta}(n, t, b, m) = \max [ x_{\delta}(n, t, m), b + p - b_{\max} ]
\end{equation*}
Let $y(n,t,b,m)$ be the rebuffering time.
\begin{equation*}
  y(n, t, b, m) = \max [ x'_{\delta}(n, t, b, m) - b, 0 ]
\end{equation*}
We generate entries for $r(n,\cdot,\cdot)$ from $r(n-1,\cdot,\cdot)$ using
\begin{equation*}
  r(n, t, b) = \max_{m, t', b'} \Big( r(n-1, t', b') + \upsilon_m - \gamma y(n, t', b', m) \Big)
\end{equation*}
such that $t = t' + x'_{\delta}(n,t',b',m)$ and \\
$b = b' - x'_{\delta}(n,t',b',m) + y(n,t',b',m) + p$.

\begin{figure}
\hrule
\begin{algorithmic}[1]
  \footnotesize
  \State $r(0, t, b) \gets \{0$ for $t = b = 0$, $-\infty$ otherwise$\}$
  \For {$n$ in $[1, N]$}
    \State initialize $r(n, t, b) \gets -\infty$ for all $t, b$
    \For {all $(t', b')$ such that $r(n-1, t', b') > -\infty$}
      \For {$m$ in $[1, M]$}
        \State $x \gets \text{ download time}(n, t', m)$
        \State $x_{\delta} \gets \lfloor x / \delta \rfloor \cdot \delta$
        \State $x'_{\delta} \gets \max[x_{\delta}, b' + p - b_{\max}]$
        \State $y \gets \max[x'_{\delta} - b', 0]$
        \State $t \gets t' + x'_{\delta}$
        \State $b \gets b' - x'_{\delta} + y + p$
        \State $r' \gets r(n-1, t', b') + \upsilon_m - \gamma y$
        \State $r(n, t, b) \gets \max[r(n, t, b), r']$
      \EndFor
    \EndFor
  \EndFor
  \State $\displaystyle r^* \gets \max_{(t,b)} \frac{r(N, t, b)}{(t + b)}$
\end{algorithmic}
\hrule
\caption{Calculating the Offline Optimal Utility Upper Bound}
  \label{alg:offline}
\end{figure}

The dynamic programming algorithm is shown in Fig.~\ref{alg:offline}.

\subsection{Evaluating BOLA-BASIC}
\label{subsec:bolaBasic}

\begin{figure}
  \centering
  \includegraphics{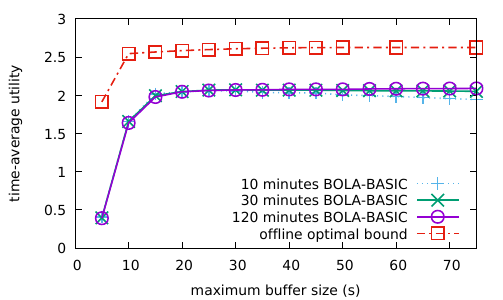}
 \caption{Time-average utility for $\gamma p =5$ using profile 1 for \mbox{BOLA-BASIC}.}
  \label{fig:BolaBasic}
\end{figure}

Fig.~\ref{fig:BolaBasic} shows the time-average utility of \mbox{BOLA-BASIC} when the video length is 10, 30 and 120 minutes. We set $\gamma p = 5$ and varied $V$ for different buffer sizes. We compared the utility of \mbox{BOLA-BASIC} with the offline optimal bound described in Section~\ref{subsec:bound}. The offline optimal gave nearly the same utility for the different video lengths. \mbox{BOLA-BASIC} only obtains about 80\% of the offline optimal bound. Also, the utility of \mbox{BOLA-BASIC} decreases slightly when the buffer size is increased because it must download more lower-bitrate segments during startup before it can reach the buffer levels required to switch to higher-bitrate segments. Our results suggests that there is room to improve \mbox{BOLA-BASIC} that motivates our next version.

\subsection{Adapting BOLA to Finite-Sized Videos}
\label{subsec:bolaFinite}

\mbox{BOLA-BASIC} was derived under the assumption that the videos are infinite. Thus, some adaptations are needed for \mbox{BOLA} to work effectively with smaller videos. Motivated by our initial experiments, we implemented two adaptations to \mbox{BOLA-BASIC} to derive a version we call \mbox{BOLA-FINITE}.

\begin{figure}
  \hrule
  \begin{algorithmic}[1]
    \footnotesize
    \For {$n$ in $[1, N]$}
    \State $t \gets \min[\text{playtime from begin}, \text{playtime to end}]$
    \State $t' \gets \max[t/2, 3p]$
    \State $Q_{\max}^{\mathrm{D}} \gets \min[Q_{\max}, t' / p]$
    \State $V^{\mathrm{D}} \gets (Q_{\max}^{\mathrm{D}} - 1) / (\upsilon_M + \gamma p)$
    \State $\displaystyle m^*[n] \gets \argmax_{m} (V^{\mathrm{D}} \upsilon_{m} + V^{\mathrm{D}} \gamma p - Q) / S_{m}$
    \If {$m^*[n] > m^*[n-1]$}
      \State $r \gets \text{bandwidth measured when downloading segment $(n-1)$}$
      \State $m' \gets \text{$\max m$ such that $S_m/p \leq \max[r,S_1/p]$}$
      \If {$m' \geq m^*[n]$}
        \State $m' \gets m^*[n]$
      \ElsIf {$m' < m^*[n-1]$}
        \State $m' \gets m^*[n-1]$
      \ElsIf {some utility sacrificed for fewer oscillations}
        \State \hspace{-1em} $\begin{array}{rl}
          \text{pause until}\hspace{-1em} & (V^{\mathrm{D}} \upsilon_{m'} + V^{\mathrm{D}} \gamma p - Q) / S_{m'} \geq \\
          & (V^{\mathrm{D}}\! \upsilon_{m'\!+1} + V^{\mathrm{D}}\! \gamma p - Q) / S_{m'\!+1}
        \end{array}$ \Comment{\mbox{BOLA-O}}
      \Else
        \State $m' \gets  m' + 1$ \Comment{\mbox{BOLA-U}}
      \EndIf
      \State $m^*[n] \gets m'$  
    \EndIf
    \State pause for $\max [p \cdot (Q - Q_{ \max}^{\mathrm{D}} + 1), 0]$
    \State download segment $n$ at bitrate index $m^*[n]$, possibly abandoning
  \EndFor
\end{algorithmic}
\hrule
\caption{The \mbox{BOLA} Algorithm.}
  \label{alg:mainalgo}
\end{figure}

1) Dynamic $V$ value for startup and wind down:
A large buffer allows \mbox{BOLA-BASIC} to perform better but it has two drawbacks. First, it takes longer to prime a large buffer during startup. Lower bitrate segments are preferred until the buffer level reaches steady state. Second, at some late stage all downloads are complete and any remaining buffered video is played out. Any available bandwidth during this period is not utilized. Shortening this period would result in less unutilized available bandwidth. We mitigate these effects by introducing a dynamic $V^{\mathrm{D}}$ which corresponds to a dynamic buffer size $Q_{\max}^{\mathrm{D}}$, shown in lines 2--5 in Fig.~\ref{alg:mainalgo}. \mbox{BOLA-FINITE} does not try to fill the whole buffer too soon and does not try to maintain a full buffer too long. We still need a minimum buffer size $3p$ for the algorithm to work effectively.

\begin{figure}
\hrule
\begin{algorithmic}[1]
  \footnotesize
  \Function{ShallAbandon}{$m, S_m^{\mathrm{R}}$}
    \State $r_m \gets (V^{\mathrm{D}} \upsilon_m + V^{\mathrm{D}} \gamma p - Q) / S_m^{\mathrm{R}}$
    \State $r_{m'} \gets (V^{\mathrm{D}} \upsilon_{m'} + V^{\mathrm{D}} \gamma p - Q) / S_{m'}$
    \State \textbf{return} \textit{true} if $r_{m'} > r_m$ for some $m'$ subject to $1 \leq m' < m$
  \EndFunction
\end{algorithmic}
\hrule
  \caption{\mbox{BOLA-FINITE}'s Download Abandonment Heuristic: $m$ is the current segment bitrate and $S_m^{\mathrm{R}}$ is the number of bits remaining to download in the current segment.}
  \label{alg:abandon}
\end{figure}

\begin{figure}
  \centering
  \includegraphics{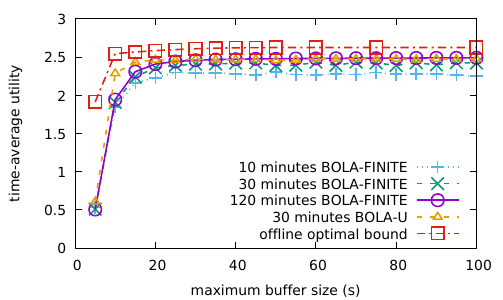}
 \caption{Time-average utility for $\gamma p =5$ using profile 1 for \mbox{BOLA-FINITE} and \mbox{BOLA-U}.}
  \label{fig:BolaFinite}
\end{figure}

\begin{figure*}
  \centering
  \includegraphics{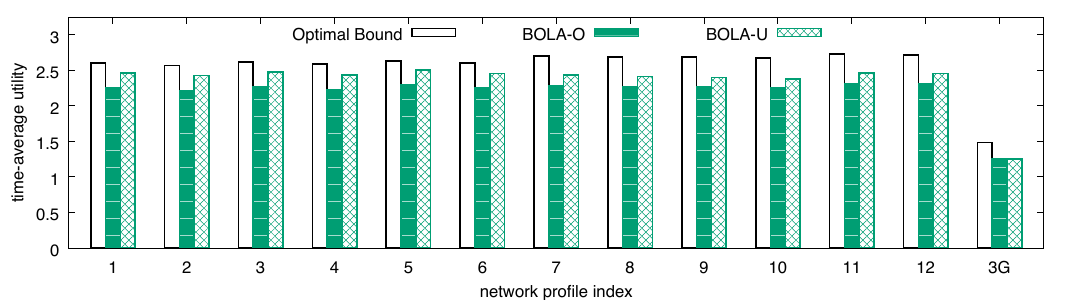}
  \caption{The time-average utility of \mbox{BOLA-O} and \mbox{BOLA-U} with $\gamma p = 5$ and a 25-second buffer playing a 30-minute video for the DASH test network profiles 1--12 and mobile traces (3G). \mbox{BOLA} utility is within 84--95\% of offline optimal utility.}
  \label{fig:UtilityA}
\end{figure*}

\begin{figure*}
  \centering
  \includegraphics{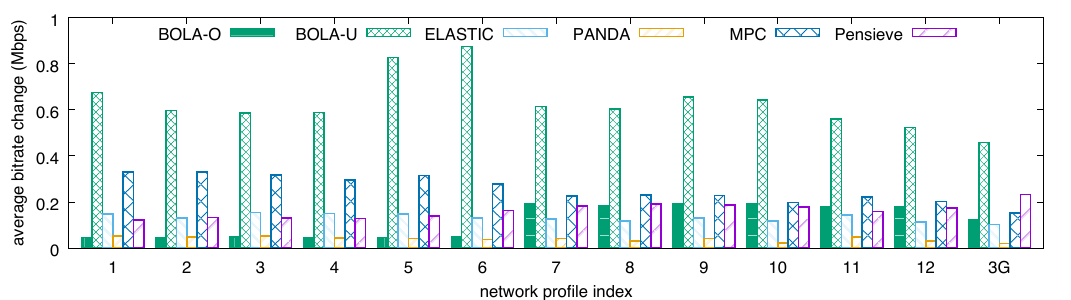}
  \caption{The average bitrate change between adjacent segments was smaller for BOLA-O than for BOLA-U, but some bitrate change is needed to accurately track the network bandwidth. In our experiments,  as an average across network profiles, ELASTIC and PANDA  tracked the bandwidth with similar accuracy to BOLA-O, while MPC and Pensieve had more oscillations.}
  \label{fig:BitrateChange}
\end{figure*}

2) Download abandonment:
\mbox{BOLA-BASIC} takes control decisions just before the download of each segment. Consider a scenario where the player is downloading high-bitrate 6~Mbps segments in good network conditions. The network bandwidth suddenly drops to 1~Mbps as the player has just started a new segment download. The segment will take $6p$~seconds to download, depleting the buffer and possibly causing rebuffering. \mbox{BOLA-FINITE} mitigates this problem by monitoring download progress and possibly abandoning a download. Fig.~\ref{alg:abandon} shows how \mbox{BOLA-FINITE} decides whether or not to abandon the download. If a segment at bitrate index $m$ is being downloaded, the remaining size $S_m^{\mathrm{R}}$ is less than $S_m$. The segment can be abandoned and downloaded at some bitrate index $m'$ subject to $1 \leq m' < m$ when $(V^{\mathrm{D}} \upsilon_m + V^{\mathrm{D}} \gamma p - Q)/S_m^{\mathrm{R}} < (V^{\mathrm{D}} \upsilon_{m'} + V^{\mathrm{D}} \gamma p - Q)/S_{m'}$. The control idea remains the same, but the current bitrate $m$ has a smaller corresponding size $S_m^{\mathrm{R}}$ because part of the segment has already been downloaded. Fig.~\ref{fig:Demo} illustrates a scenario where abandonment might help. At 46s a 3~Mbps segment download starts. Since there is a bandwidth drop at the time, the segment takes almost 9s to download. The buffer is depleted and \mbox{BOLA-BASIC} switches to downloading at a bitrate of 0.3~Mbps. \mbox{BOLA-FINITE} with abandonment logic would have detected the rapidly depleting buffer and stopped the long download, with the system only dropping to the 1.4 and 0.7~Mbps download bitrates in the low-bandwidth period.

Fig.~\ref{fig:BolaFinite} shows the time-average utility of \mbox{BOLA-FINITE} for 10, 30 and 120 minutes of playback time with $\gamma p=5$. Comparing with \mbox{BOLA-BASIC} in Fig.~\ref{fig:BolaBasic}, we see that the time-average utility is much closer to the offline optimal bound. The benefit of the adjustments is also evident as the buffer grows larger, as there is no significant decrease in utility caused by filling the buffer with low-bitrate segments in the earlier stages of the video.

\subsection{Avoiding Bitrate Oscillations}
\label{subsec:oscillations}

While our performance objective optimizes playback utility and playback smoothness, users are also sensitive to excessive bitrate switching. We discuss three causes of bitrate switches.

1) Bandwidth variation:
As the network conditions change, the player varies the bitrate, tracking the network bandwidth. Such switches are acceptable; the player has no control on the bandwidth and should adapt to different network conditions.

2) Dense buffer thresholds:
Either a larger number of bitrate levels and/or a smaller buffer size may push the threshold levels closer. If the differences between threshold levels are less than the segment duration $p$, adding one downloaded segment to the buffer may push the buffer level over several threshold levels at once. This might cause \mbox{BOLA-FINITE} to overshoot and choose a bitrate that is too high for the available bandwidth. Consequently, the segment download would take much more than $p$ seconds, leading to excessive buffer depletion, causing \mbox{BOLA-FINITE} to switch down its bitrate by more than one level. In such a scenario \mbox{BOLA-FINITE} can oscillate between bitrates, even when the available bandwidth is stable.

3) Bitrate quantization:
Having a stable network bandwidth and widely-spaced thresholds still does not avoid all bitrate switching. Suppose the bandwidth is 2.0~Mbps and it lies between two encoded bitrates of 1.5 and 3.0~Mbps. While the player downloads 1.5~Mbps segments, the buffer keeps growing. When the buffer crosses the threshold the player switches to 3.0~Mbps, depleting the buffer. After the buffer gets sufficiently depleted, the player switches back to 1.5~Mbps, and the cycle repeats. In this example, a viewer might prefer the video player to stick to the 1.5~Mbps bitrate, sacrificing some utility in order to have fewer oscillations. Or, a viewer might want to maximize utility and play a part of the video in the higher bitrate of 3.0~Mbps at the cost of more oscillations. We describe two variants of \mbox{BOLA} below to suit either viewer.

\begin{figure*}
  \centering
  \includegraphics{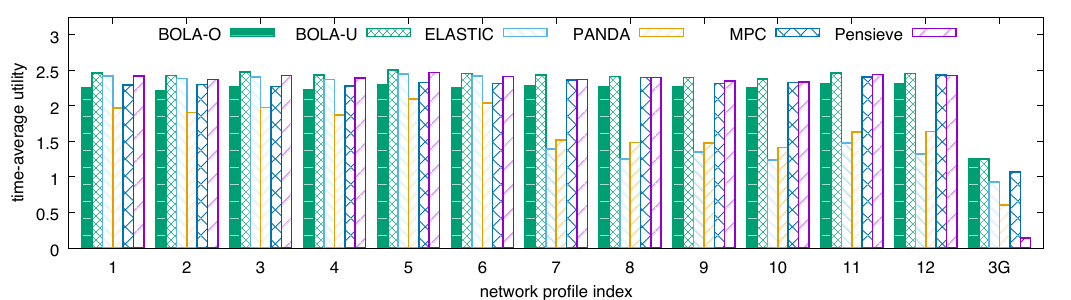}
  \caption{The time-average utility of \mbox{BOLA-O}, \mbox{BOLA-U}, ELASTIC, PANDA, MPC and Pensieve with $\gamma p = 5$ playing a 30-minute video for the DASH test network profiles 1--12 and mobile traces (3G). Compared with ELASTIC and PANDA, \mbox{BOLA-U} has  about 1.75 times the utility of the other algorithms in roughly half the cases. MPC has a utility between \mbox{BOLA-O} and \mbox{BOLA-U}. Pensieve has a utility between BOLA-O and BOLA-U for profiles 1--12 but performs worse for the mobile (3G) traces.}
  \label{fig:UtilityB}
\end{figure*}

\begin{figure*}
  \centering
  \includegraphics{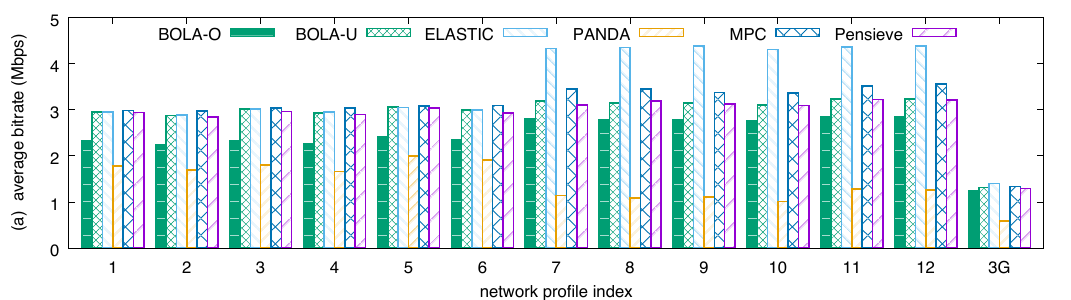} \\
  \includegraphics{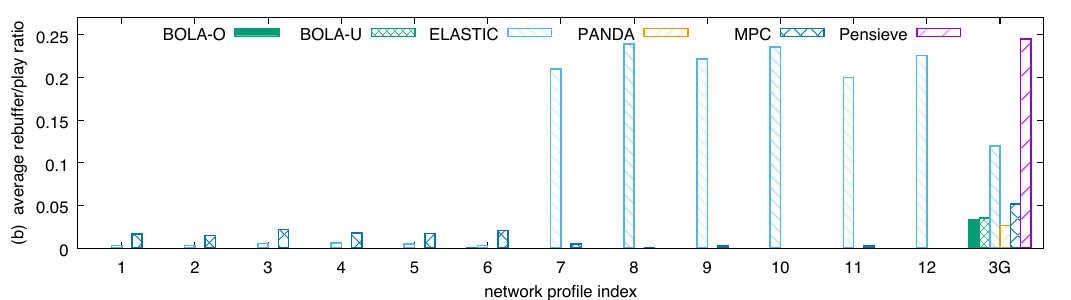}
  \caption{Comparing \mbox{BOLA} with ELASTIC, PANDA, MPC and Pensieve using raw metrics: average bitrate and rebuffer-to-play ratio. \mbox{BOLA}, PANDA and Pensieve do not rebuffer for profiles 1--12. ELASTIC has almost no rebuffering for profiles 1--6, but it has a rebuffer-to-play ratio greater than 20\% for profiles 7--12. MPC has some rebuffering for almost all profiles. Pensieve has no rebuffering for profiles 1--12. But, Penseive  has a 24\% rebuffer-to-play ratio for the mobile (3G) traces, as it is unable to perform well for bandwidth conditions that are significantly different from its training set.}
  \label{fig:AverageBitrateRebuffer}
\end{figure*}

The first variant that we call \mbox{BOLA-O} mitigates \textbf{\underline{o}}scillations by introducing bitrate capping (lines 7--20 in Fig.~\ref{alg:mainalgo}) when switching to a higher bitrate. \mbox{BOLA-O} verifies that the higher bitrate is sustainable by comparing it to the bandwidth as measured when downloading the previous segment (lines 8--11). Since the motive is to limit oscillations rather than to predict future bandwidth, this adaptation does not drop the bitrate to a lower level than in the previous download (lines 12--13). Continuous downloading at a bitrate lower than the bandwidth would cause the buffer to keep growing. \mbox{BOLA-O} avoids this by allowing the buffer to slip to the appropriate threshold before starting the download (line 15).

The second variant that we call \mbox{BOLA-U} does not sacrifice \textbf{\underline{u}}tility. Excessive buffer growth is avoided by allowing the bitrate to be one level higher than the sustainable bandwidth (line 17). This allows the player to choose 3~Mbps in the example. While \mbox{BOLA-U} does not handle the third type of oscillations, it handles the more severe second type.

Looking back at Fig.~\ref{fig:BolaFinite}, we see that the added stability of \mbox{BOLA-U} pays off when using a small buffer size and BOLA-U achieves a larger utility than BOLA-FINITE. Fig.~\ref{fig:UtilityA} shows the time-average utility of \mbox{BOLA-O} and \mbox{BOLA-U} with $\gamma p = 5$ and $Q_{\max}p = 25$s playing a 30-minute video. The utility lost by \mbox{BOLA-O} to avoid oscillations is clearly evident. In practice the lost utility is limited by the distance between encoded bitrates; if the next lower bitrate level is not far from the network bandwidth, then little utility will be lost.

We measure oscillations by comparing consecutive segments. The change in bitrate between a segment and the next is the absolute difference between bitrates (in Mbps) of the two segments. Fig.~\ref{fig:BitrateChange} shows the bitrate change averaged across all the segments. While \mbox{BOLA-U} has a high average bitrate change because of the quantization, \mbox{BOLA-O} only switches bitrate because of network bandwidth variations.

\begin{figure*}
  \centering
  \includegraphics{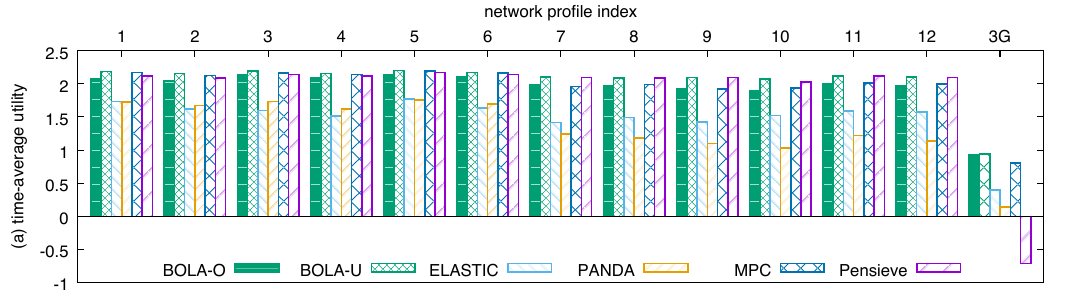} \\[-6pt]
  \includegraphics{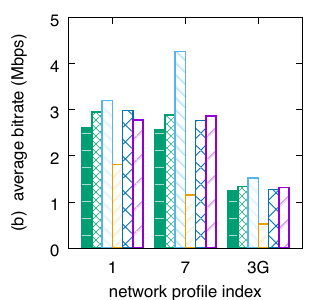} \hfill
  \includegraphics{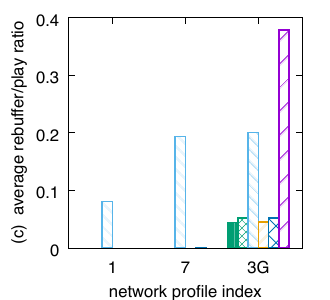} \hfill
  \includegraphics{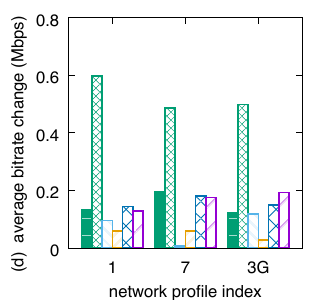}
  \caption{The time-average utility of BOLA-O, BOLA-U, ELASTIC, PANDA, MPC and Pensieve with $\gamma p = 5$ playing a different video for 30 minutes, using the DASH test network profiles 1--12 and the mobile traces (3G). Note that Pensieve has negative 3G utility because of excessive rebuffering (the average rebuffer-to-play ratio is 38\%). The raw metrics are also provided in the plots above.}
  \label{fig:UtilityOther}
\end{figure*}

\subsection{Comparison With State-of-the-Art Algorithms}

\label{sec:CompareAlgorithms}

We now compare \mbox{BOLA} with four state-of-the art algorithms, ELASTIC \cite{decicco2013elastic}, PANDA \cite{li2014panda}, MPC \cite{yin2015ccontrol} and Pensieve~\cite{mao2017neural}.
We use the default design parameters in \cite{decicco2013elastic,li2014panda,yin2015ccontrol,mao2017neural}. We test both \mbox{BOLA-O} and \mbox{BOLA-U}. Although \mbox{BOLA} performs better with larger buffers, we limited the buffer size to 25s for the tests to ensure fairness. ELASTIC targets a buffer level of 15s but the buffer level varies higher. PANDA targets a minimum buffer level of 26s. We use the RobustMPC variant of MPC with a buffer size of 25s. MPC relies on bandwidth estimation; we use the harmonic mean over the last five segment downloads to be consistent with the empirical evaluation method in \cite{yin2015ccontrol}. We trained a Pensieve neural network model for the video with a buffer size of 25s. For  training Pensieve, we used bandwidth traces generated using the tool provided in the Pensieve repository as recommended.

Fig.~\ref{fig:UtilityB} compares the algorithms using each of the 12 network profiles and the mobile traces. \mbox{BOLA-U} consistently performs significantly better than PANDA. While \mbox{BOLA-U} and ELASTIC perform similarly for profiles 1--6, \mbox{BOLA-U} performs significantly better for the other profiles that have larger bandwidth variations. MPC and Pensieve consistently obtains a utility between \mbox{BOLA-O} and \mbox{BOLA-U} for profiles 1--12, but perform worse for the mobile traces.
We repeat the comparison using the average bitrate and rebuffering metrics in Fig.~\ref{fig:AverageBitrateRebuffer}. This gives an insight into the strengths and weaknesses of the different algorithms.

\textbf{Comparing BOLA-U with ELASTIC:} For profiles 1--6, \mbox{BOLA-U} has approximately the same bitrate as ELASTIC. ELASTIC has a higher bitrate for profiles 7--12, but that comes at a significant cost in terms of rebuffering. For these profiles, the ratio of the rebuffering time to the play time is more than 20\% for ELASTIC, while \mbox{BOLA-U} has no rebuffering. For the mobile traces, ELASTIC has marginally higher bitrate than \mbox{BOLA-U} but has a 12.0\% rebuffer-to-play ratio compared with \mbox{BOLA-U}'s 3.5\%. ELASTIC rebuffers significantly more because it does not react in time when the bandwidth drops. 

\textbf{Comparing BOLA-U with PANDA:} Both algorithms do not rebuffer for profiles 1--12.  For the mobile traces, \mbox{BOLA-U}  and PANDA have a rebuffer-to-play ratio of 3.5\% and 2.6\% respectively. However, PANDA has significantly lower bitrate than \mbox{BOLA-U}. The reason is that PANDA is more conservative and in some cases does not change to a higher bitrate even if it is sustainable.

\textbf{Comparing BOLA-U with MPC:} Both algorithms have similar average bitrates but MPC has slightly higher bitrate for some of the profiles. However, while \mbox{BOLA-U} does not rebuffer, MPC has some rebuffering for most of the profiles. While it is possible to tune the MPC parameters to avoid that rebuffering, it is not clear how to choose parameters that consistently work for different network conditions. Another factor that might contribute to MPC rebuffering is the bandwidth estimation. When there is a large drop in bandwidth, the recommended harmonic mean bandwidth estimator takes a while to react. Even though RobustMPC factors in network estimation error, rebuffering is not totally eliminated.

\textbf{Comparing BOLA-U with Pensieve:} For profiles 1--12, Pensieve obtains utility between BOLA-O and BOLA-U, but consistently closer to  BOLA-U. However, Pensieve has too much rebuffering in the mobile traces, resulting in much worse utility for these traces. While the network traces used to train Pensieve included periods with low bandwidth similar to the mobile traces, Pensieve did not learn a model that would perform well in relatively low bandwidth situations in a mobile setting. This points to a weakness in Pensieve as it is unable to adapt to bandwidth conditions that are significantly different from the training set.

In Fig.~\ref{fig:BitrateChange} we show our results for our secondary metric of bitrate oscillations. \mbox{BOLA-U} does not perform well in this metric, since it attempts to maximize utility at the cost of increased oscillations. Comparing \mbox{BOLA-O} with ELASTIC, PANDA, MPC, and Pensieve, ELASTIC has a lower average change than \mbox{BOLA-O} only in the cases where it has a slow reaction and excessive rebuffering. PANDA has a lower average change because it is more conservative and in some cases does not change to a higher bitrate even if that bitrate is sustainable. MPC has higher average change than \mbox{BOLA-O} for profiles 1--12. Pensieve has similar average change to \mbox{BOLA-O} for profiles 1--12.

We also tested the algorithms with more videos to investigate performance when changing characteristics such as content type, segment duration, and available bitrates. The tests showed similar results. One example is the video provided with Pensieve. The video has 49 segments with a segment duration of 4s. It is encoded at six bitrates: 0.3, 0.75, 1.2, 1.85, 2.85 and 4.3~Mbps. Fig.~\ref{fig:UtilityOther} shows the utility and metrics for the same six algorithms with similar conclusions to Figs.~\ref{fig:BitrateChange}--\ref{fig:AverageBitrateRebuffer}. Note that Pensieve fails to perform well on mobile traces again, since it is significantly different from its training set.

{\em Thus, from our empirical analysis, we can conclude that \mbox{BOLA} achieves higher utility, and performs more consistently across different scenarios in comparison with ELASTIC, PANDA, MPC and Pensieve.} One reason for the consistency of \mbox{BOLA} is that it does not have a large number of parameters. \mbox{BOLA} has two design parameters $\gamma$ and $V$, which have an intuitive significance as discussed in Section~\ref{subsec:param}, and an option of whether or not to trade off some utility to reduce oscillations. Other algorithms have a number of different parameters and tuning the parameters for a particular scenario might make the system less suited for other scenarios. Also, BOLA's ability to abandon a segment during a download and start the download at a lower bitrate allows BOLA to achieve significantly less rebuffering than the other algorithms.

{\bf Note:} A number of recent papers compare new algorithms with BOLA. Some of the prior work used the experimental version of BOLA in dash.js versions 2.0.0--2.5.0 that required bug fixes. We suggest that a stable version used in production (dash.js version 2.6.0 or later) be used for such comparisons for a more accurate evaluation of BOLA.  Also, dash.js has a default buffer size of 12s, leading some researchers to compare a large-buffer algorithm with a small-buffer BOLA. Our work uses the correct BOLA implementation and the same buffer size for all the compared algorithms.

%% file: deployment.tex
\section{Deployment}

\label{sec:deployment}

\subsection{The DASH Reference Player}

After developing a theoretical foundation for BOLA and testing it by simulation, we deployed BOLA in a production setting. Particularly, we implemented BOLA in dash.js, the open-source standard DASH reference player~\cite{dashjs}. Through dash.js, BOLA is now being used in production by several major video providers and delivery networks such as Akamai, BBC, CBS and Orange. Deployment in production presented a number of new challenges such as operating with even smaller buffer capacities, correctly handling events such as a user seek to a different point in the video, and tolerating delays caused by the video player unrelated to the network conditions. The techniques we implemented to handle these new challenges are described in~\cite{Spiteri2018}.

\subsection{BOLA Parameters}

\label{sec:deployparam}

One deployment challenge involves choosing the BOLA parameters $\gamma$ and $V$. We gave an intuition to pick the parameters in Section~\ref{subsec:param}, but video providers are more familiar with choosing buffer level targets. For this purpose,
we now discuss how to derive $\gamma$ and $V$ from intuitive requirements. Consider the following requirements:\\
1. We want a maximum buffer level $Q_{\max}$.\\
2. We want to download at the highest bitrate when the buffer level is $Q_{\max}$.\\
3. We want to download at the lowest bitrate when the buffer level is less than a threshold $Q_{\mathrm{low}}$,  and we want to download at a higher bitrate when the buffer level goes above the threshold.\\
These requirements are easy to understand for video providers who might not be familiar with \mbox{BOLA}. In fact, video providers usually have some preferred maximum buffer level $Q_{\max}$. Further, they might have preference for $Q_{\mathrm{low}}$ such as 10s as described in \cite{Spiteri2018}.

To satisfy requirements 1--2, we want \eqref{eq:lyp_algo} to switch from choosing $a_M = 1$ to choosing $\sum a_m=0$ at the threshold when the buffer level is $Q_{\max}$. This happens if
\begin{align}
  \rho_{a_M=1} &= \rho_{\mathbf{a}=\mathbf{0}} \nonumber \\[6pt]
  \frac{V(\upsilon_M + \gamma p) - Q_{\max}}{S_M} &= 0
  \label{eq:Vg1}
\end{align}
Note that BOLA satisfies requirement 2 and downloads at the highest bitrate just before the $Q_{\max}$ threshold because at that buffer level we get $\rho_{a_m=1} < \rho_{a_M=1}$ for $m < M$. This is illustrated in Fig.~\ref{fig:DemoThreshold}.

To satisfy requirement 3, we want \eqref{eq:lyp_algo} to switch from choosing $a_1 = 1$ to choosing $a_2=1$ at the threshold when the buffer level is $Q_{\mathrm{low}}$. This happens if
%
\begin{align}
  \rho_{a_1=1} &= \rho_{a_2=1} \nonumber \\[6pt]
  \frac{V(\upsilon_1 + \gamma p) - Q_{\mathrm{low}}}{S_1} &=
  \frac{V(\upsilon_2 + \gamma p) - Q_{\mathrm{low}}}{S_2}
  \label{eq:Vg2}
\end{align}

Solving \eqref{eq:Vg1}--\eqref{eq:Vg2}, we obtain
\[
V = \frac{Q_{\max} - Q_{\mathrm{low}}}{\upsilon_M - \alpha}, \quad
\gamma p = \frac{\upsilon_M Q_{\mathrm{low}} - \alpha Q_{\max}}{Q_{\max} - Q_{\mathrm{low}}}
\]
  where
\[
\alpha = \frac{S_1\upsilon_2 - S_2\upsilon_1}{S_2 - S_1}.
\]

When calculating the BOLA parameters from $Q_{\mathrm{low}}$ and $Q_{\max}$, the previous intuition about $\gamma$ and $V$ still hold. If a video provider chooses a larger $Q_{\mathrm{low}}$, $\gamma$ will be larger and BOLA will give more weight to rebuffering. If a video provider chooses a larger $Q_{\max}$, $V$ will be larger.

%% file: related.tex
\section{Related Work}
\label{sec:related}

There has been a lot of recent work on bitrate adaptation algorithms, much of which is based on estimating the bandwidth of the network connection.
FESTIVE \cite{jiang2014festive} uses a harmonic bandwidth estimator to predict future bandwidth from past downloads, limiting bitrate change to one level between successive segments for stability. Notably, FESTIVE attempts to find a tradeoff between efficiency and fairness with competing downloads.
BBA \cite{Huang+14} is a buffer-based algorithm. BOLA has a few similarities to BBA but the mapping function from buffer level to video bitrate is different. Also, BBA assumes that the buffer size is large (in the order of minutes), thereby making it not suitable for short videos. Further, it does not provide any theoretical guarantees for its buffer-based approach.
A notable algorithm is ELASTIC \cite{decicco2013elastic} that uses control theory to adjust the bitrate so as to keep the buffer occupancy at a constant level.
Another notable algorithm is PANDA \cite{li2014panda} which also estimates the network bandwidth. PANDA drops the download bitrate as soon as low bandwidth is detected but only increases the bitrate slowly to probe the real capacity when a higher bandwidth is detected. Like FESTIVE, PANDA trades efficiency for fairness.
In \cite{yin2015ccontrol}, an algorithm using model predictive control (MPC) is proposed to optimize a comprehensive set of metrics. In this approach, the bitrate for the current segment is chosen based on a network  bandwidth prediction for the next few segments. But, its performance depends on the accuracy of such a prediction. The approach also requires significant offline optimization to be performed outside of the client for an exhaustive set of scenarios.
\cite{mao2017neural} presents Pensieve, a reinforcement-learning approach to ABR. A neural network model can be trained for a video using a particular buffer size, using a QoE function for reward. A set of bandwidth traces is used as training data. Unfortunately, a trained model does not transfer easily to a different video or, more importantly, to bandwidth conditions not represented in the training data.
Unlike prior work, we derive a buffer-based algorithm with theoretical guarantees that is simple to implement within the client and we empirically show its efficacy on extensive network traces. In recent work \cite{akhtar2018oboe}, a method called Oboe for auto-tuning the parameters of BOLA and MPC was presented and shown to improve both algorithms. Further, the work showed that Oboe used in conjunction with traditional ABR algorithms performs better than reinforcement-learning based ABR such as Pensieve.

%% file: conclusion.tex
\section{Conclusion}
\label{sec:conclusion}

We formulated video bitrate adaptation for ABR streaming as a utility maximization problem and derived \mbox{BOLA}, an online control algorithm that is provably near-optimal. Further, we empirically demonstrated the efficacy of \mbox{BOLA} using extensive traces. In particular, we showed that our online algorithm achieves utility close to the optimal offline algorithm. We showed that our algorithm
performs better than state-of-the-art algorithms in a number of different test scenarios.
We also implemented BOLA in dash.js, the open-source standard DASH reference player \cite{dashjs}. Through dash.js, BOLA is now being used in production by several major video providers and delivery networks such as Akamai, BBC, CBS and Orange.

%% file: appendix.tex
\appendix[\texorpdfstring{Proof of Theorem~\ref{thm:BOLA}}{Proof or Theorem 1}]
\label{sec:NP-proof}

We first show part $1$ using induction. Note that the bound $Q(t_k) \leq V(\upsilon_M + \gamma p) + 1$ holds for $k=1$ since $Q(t_1) = Q(0) = 0$. Now suppose it holds for some $k$. We will show that it will also hold for $k+1$. We have two cases.

\emph{Case 1:} $Q(t_k) \leq V(\upsilon_M + \gamma p)$ \\
From the queueing equation (\ref{eq:Q}), it follows that the maximum that $Q(t_k)$ can increase in slot $k$ is by $1$. This implies that $Q(t_{k+1}) \leq V(\upsilon_M + \gamma p) + 1$.

\emph{Case 2:} $V(\upsilon_M + \gamma p) < Q(t_k) \leq V(\upsilon_M + \gamma p) + 1$ \\
We have $Q(t_k) > V ( \upsilon_m + \gamma p)$ for all $m \in \{1, 2, \ldots, M\}$ (using (\ref{eq:upsilon})). It follows from the structure of optimal solution to (\ref{eq:lyp_algo}) that \mbox{BOLA} will choose the no-download option in this case. As a result, $Q(t_{k})$ cannot increase and we have that $Q(t_{k+1}) \leq V(\upsilon_M + \gamma p) + 1$.

$Q(t_k)$ denotes the total number of segments in the buffer. This can be at most $Q_{\max}$ using the relation
\begin{equation*}
  V \leq \frac{Q_{\max} - 1}{\upsilon_M + \gamma p}.
\end{equation*}

In part 2, we show the bound in (\ref{eq:bound2}) using the technique of Lyapounov optimization over variable size frames \cite{neely2010}. We first define a Lyapunov function $L(Q(t_k))$ as
\begin{equation*}
  L(Q(t_k)) = \frac{1}{2} Q^2(t_k)
\end{equation*}
and define the per-slot conditional Lyapunov drift $D(t_k)$ as
\begin{equation*}
  D(t_k) \defequiv \expect{L(Q(t_{k+1})) - L(Q(t_k)) | Q(t_k)}.
\end{equation*}
We use the queueing equation \eqref{eq:Q}, to bound $D(t_k)$. We consider two cases for \eqref{eq:Q}: $Q(t_k) \leq T_k/p$ and $Q(t_k) > T_k/p$. In the first case we have
\begin{equation*}
  D(t_k) = \mathbb{E} \Bigg\{ \frac{1}{2}\left(\sum_{m=1}^M a_m(t_k)\right)^2 - \frac{1}{2}Q^2(t_k) | Q(t_k) \Bigg\}.
\end{equation*}
In the second case we have
\begin{multline*}
  D(t_k) = \mathbb{E} \Bigg\{\frac{1}{2}\left(\frac{T_k}{p} - \sum_{m=1}^M a_m(t_k) \right)^2 \\
  - Q(t_k) \left(\frac{T_k}{p} - \sum_{m=1}^M a_m(t_k) \right) | Q(t_k) \Bigg\}
\end{multline*}
In both cases, $D(t_k)$ is bounded by
\begin{align}
D(t_k) \leq \frac{p^2 + \Psi}{2p^2} - Q(t_k) \expect{\frac{T_k}{p} - \sum_{m=1}^M a_m(t_k) | Q(t_k)}
\label{eq:bound_drift}
\end{align}
where $\Psi$ is an upper bound on $\expect{T_k^2}$ under any control algorithm and is assumed to be finite.

Following the methodology of the Lyapunov optimization technique, we subtract $V \times \textrm{reward term}$ from both sides of the above to get
\begin{align}
D(t_k) &- V \expect{\sum_{m=1}^M a_m(t_k)(\upsilon_m + \gamma p) |  Q(t_k)}  \nonumber\\
& \leq \frac{p^2 + \Psi}{2p^2} - Q(t_k) \expect{\frac{T_k}{p} - \sum_{m=1}^M a_m(t_k) | Q(t_k)} \nonumber\\
&- V \expect{\sum_{m=1}^M a_m(t_k)(\upsilon_m + \gamma p) |  Q(t_k)}
\label{eq:drift1}
\end{align}
Let us denote the control decisions (and resulting slot lengths) under our control algorithm by the superscript \mbox{BOLA} while those under the stationary policy of Lemma~\ref{lem:policy} by STAT. Since \mbox{BOLA} greedily maximizes over a frame, it ensures that
\begin{multline}
{\expect{\sum_{m=1}^M a_m^{\mathrm{BOLA}}(t_k) (Q(t_k) - V (\upsilon_m + \gamma p))  | Q(t_k)}}  \\
\leq  \eta \times {\expect{\sum_{m=1}^M a_m^{\mathrm{STAT}}(t_k) (Q(t_k) - V (\upsilon_m + \gamma p))  | Q(t_k)}}
\label{eq:bolastat}
 \end{multline}
where $\eta = \frac{\expect{T_k^{\mathrm{BOLA}} | Q(t_k)}}{\expect{T_k^{\mathrm{STAT}} | Q(t_k)}}$. To see this, compare the ratio on the left hand side above with the objective in (\ref{eq:lyp_algo}) while noting that we can express the denominator as $\expect{T_k^{\mathrm{BOLA}} | Q(t_k)} = (\sum_{m=1}^M a_m^{\mathrm{BOLA}}(t_k) S_m )/\omega_{\mathrm{avg}}$. It should be noted that this ratio can be minimized without requiring knowledge of $\omega_{\mathrm{avg}}$. Then we use \eqref{eq:bolastat} to express \eqref{eq:drift1} as
\begin{align*}
&D^{\mathrm{BOLA}}(t_k) - V \expect{\sum_{m=1}^M a_m^{\mathrm{BOLA}}(t_k)(\upsilon_m + \gamma p) |  Q(t_k)}  \nonumber\\
& \leq \frac{p^2 + \Psi}{2p^2} - Q(t_k) \expect{\frac{T_k^{\mathrm{BOLA}}}{p} - \eta \sum_{m=1}^M a_m^{\mathrm{STAT}}(t_k) | Q(t_k)} \nonumber\\
  &\;\; - V \eta \expect{\sum_{m=1}^M a_m^{\mathrm{STAT}}(t_k)(\upsilon_m + \gamma p) |  Q(t_k)}\nonumber
\end{align*}
Substituting the time-average values for the stationary policy we get
\begin{align}
&D^{\mathrm{BOLA}}(t_k) - V \expect{\sum_{m=1}^M a_m^{\mathrm{BOLA}}(t_k)(\upsilon_m + \gamma p) |  Q(t_k)}  \nonumber\\
& \leq \frac{p^2 + \Psi}{2p^2} - Q(t_k) \Big(\frac{1}{p} - r^{\mathrm{STAT}}\Big) \expect{T_k^{\mathrm{BOLA}}| Q(t_k)} \nonumber\\
&\;\; - V (\upsilon^* + \gamma s^*) \expect{T_k^{\mathrm{BOLA}}| Q(t_k)}
\label{eq:drift3}
\end{align}
where $r^{\mathrm{STAT}}$ denotes the expected arrival rate under the stationary policy and cannot exceed $1/p$ since it is rate stable. Thus we have
\begin{align}
&D^{\mathrm{BOLA}}(t_k) - V \expect{\sum_{m=1}^M a_m^{\mathrm{BOLA}}(t_k)(\upsilon_m + \gamma p) |  Q(t_k)}  \nonumber\\
& \leq \frac{p^2 + \Psi}{2p^2} - V (\upsilon^* + \gamma s^*) \expect{T_k^{\mathrm{BOLA}}| Q(t_k)}
\label{eq:drift4}
\end{align}
Taking conditional expectation of both sides and summing over $k \in \{1, 2, \ldots, K_N\}$, we get
\begin{align}
&\expect{L(Q(t_{K_{N+1}}))} - V \expect{\sum_{k=1}^{K_N} \sum_{m=1}^M a_m^{\mathrm{BOLA}}(t_k)(\upsilon_m + \gamma p)}  \nonumber\\
& \leq \frac{(p^2 + \Psi)K_N}{2p^2} - V (\upsilon^* + \gamma s^*) \expect{\sum_{k=1}^{K_N} T_k^{\mathrm{BOLA}}}
\label{eq:drift5}
\end{align}
Dividing both sides by $V\expect{\sum_{k=1}^{K_N} T_k^{\mathrm{BOLA}}}$ and taking the limit as $N \to \infty$ yields the bound in (\ref{eq:bound2}).